\documentclass[prd,aps,nofootinbib,floatfix]{revtex4}
\usepackage{mathtools}
\usepackage{varwidth}
\usepackage{amsmath}
\usepackage{amssymb}
\usepackage{cancel}

\makeatletter

\usepackage{graphicx}\usepackage{epsfig}\usepackage{dsfont}
\usepackage[usenames]{color}
\usepackage{ulem}
\usepackage{bigstrut}
\usepackage{multirow}
\usepackage{subfigure}

\allowdisplaybreaks

\makeatother

\begin{document}
\title{Semileptonic and nonleptonic $\bar{B}_{s}\to D_{sJ}$ decays in covariant light-front approach}
\author{
Wuenqi$^{1}$, 
Run-Hui Li$^{1}$~\footnote{Email: lirh@imu.edu.cn}, 
Zhen-Xing Zhao$^{1}$~\footnote{Email: zhaozx19@imu.edu.cn (Corresponding author)}
}
\affiliation{$^{1}$ School of Physical Science and Technology, \\
 Inner Mongolia University, Hohhot 010021, China }

\begin{abstract}
In this work, we investigate the $\bar{B}_{s}\to D_{sJ}$ transitions in covariant light-front approach,
where $D_{sJ}$ denotes an s-wave or p-wave $c\bar{s}$ meson state.
We first obtain the transition form factors within the framework of the quark model,
and then apply the obtained form factors to obtain the branching fractions of semileptonic and nonleptonic decays.
We also compare our results with experimental data and other theoretical predictions.
We find that some branching fractions are sizable and might be accessible
at the LHC and future $e^{+}$-$e^{-}$ colliders.
Our work is expected to be of great value for establishing corresponding decay channels,
and also be helpful in understanding the non-perturbative QCD dynamics. 
\end{abstract}

\maketitle

\section{Introduction}

Meson decays not only help us to test the theoretical framework of the Standard Model (SM)
but also offer opportunities to search for new physics beyond
the SM. Among others, the $\bar{B}_{s}\to D_{sJ}$
transitions have attracted significant attention in recent years,
where $D_{sJ}$ denotes an s-wave or p-wave $c\bar{s}$ meson state.
The s-wave $c\bar{s}$ states include $D_{s}$ and $D_{s}^{*}$,
while the p-wave ones include the scalar $D_{s0}$,
the two axial-vectors $D_{s1}$ and $D_{s1}^{\prime}$, and the tensor $D_{s2}$.
Although corresponding meson states have been established experimentally,
there are still many problems to be solved, such as the mixing problem of the two axial-vectors.
Investigation of the $\bar{B}_{s}\to D_{sJ}$ transitions not only sheds light on
the internal structure of these $D_{sJ}$ states,
but also is helpful in understanding the underlying non-perturbative QCD dynamics.

In this work, we employ the covariant Light-Front Quark Model (LFQM)
to investigate the $\bar{B}_{s}\to D_{sJ}$ transitions, which are
induced at the quark level through the $b\to c$ process,
and the anti-strange quark acts as a spectator. 
The covariant LFQM offers a relativistic treatment of hadrons,
allowing us to account for their spin structure through the Melosh rotation.
The light-front wave functions, which characterize a hadron
in terms of its internal quark and gluon degrees of freedom,
are independent of the momentum of the hadron, ensuring Lorentz invariance.
Therefore, this approach provides a natural way to calculate the transition amplitudes
between different hadron states. By using the covariant LFQM, we aim to extract
the form factors, and then apply them to obtain some observables
in semileptonic and nonleptonic $\bar{B}_{s}\to D_{sJ}$ decays. 

The covariant LFQM has been successfully applied to study the decay
constants and form factors in the meson sector \cite{Jaus:1999zv,Cheng:2003sm,Li:2010bb,Verma:2011yw,Shi:2016gqt,Yang:2025zvo,Chang:2020wvs,Shi:2023qnw}.
In recent years, the light-front approach has also been widely applied to investigate
the properties of heavy baryons \cite{Wang:2017mqp,Zhao:2018zcb,Zhao:2018mrg,Xing:2018lre,Zhao:2022vfr,Xing:2024okl,Ke:2007tg,Ke:2012wa,Chua:2018lfa,Liu:2023zvh,Wang:2024mjw,Deng:2023qaf,Li:2021qod}.
Due to the presence of three quarks in baryons, the development of this theory went through
two stages: the diquark picture and the three-quark picture. Recently,
Ref. \cite{Zhao:2023yuk} re-established the three-quark picture
of baryons using a bottom-up approach, and the method of constructing
flavor-spin wave functions in the article is expected to be extended
to multi-quark states. However, it is worth noting that there is currently
no established covariant version of LFQM in the baryon sector.

The rest of this paper is organized as follows. In Sect. II, we provide
a brief description of the covariant LFQM, followed by
the calculation of decay constants and form factors in this model.
In Sect. III, we present our numerical results of form factors for
various transitions, then we use the derived form factors to study
semileptonic and nonleptonic decays, presenting predictions for branching
fractions. We conclude this paper in the last section.

\section{Theoretical framework}

\subsection{Covariant light-front approach}

\begin{figure}
\includegraphics[scale=0.5]{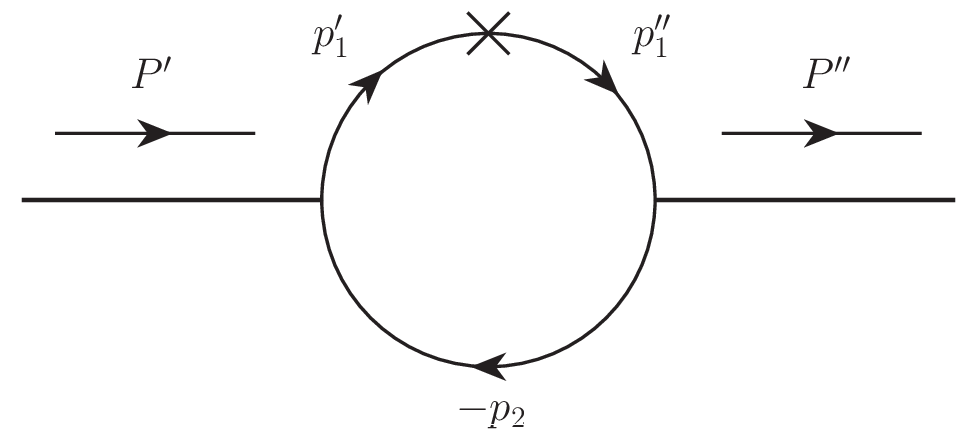}
\caption{Feynman diagram for the meson weak transitions, where the
cross ``$\times$" stands for the weak interaction.}
\label{fig:feyn}
\end{figure}

In the light-front approach, the momentum is decomposed into its light-front
form $P=(P^{-},P^{+},P_{\perp})$, where $P^{\pm}=P^{0}\pm P^{3}$
and $P_{\perp}=(P^{1},P^{2})$. For the meson weak transitions, as depicted
in Fig. \ref{fig:feyn}, the incoming (outgoing) meson has the momentum
$P^{\prime}=p_{1}^{\prime}+p_{2}$ ($P^{\prime\prime}=p_{1}^{\prime\prime}+p_{2}$)
and the mass $M^{\prime}$ ($M^{\prime\prime}$); The quark and antiquark
inside the incoming (outgoing) meson have the momenta $p_{1}^{\prime}$
($p_{1}^{\prime\prime}$) and $p_{2}$, and the masses $m_{1}^{\prime}$
($m_{1}^{\prime\prime}$) and $m_{2}$. The quark momenta can be written
in terms of the internal variables ($x_{i},p_{\perp}^{\prime}$) by:
\begin{equation}
p_{1,2}^{\prime+}=x_{1,2}P^{\prime+},\quad p_{1,2\perp}^{\prime}=x_{1,2}P_{\perp}^{\prime}\pm p_{\perp}^{\prime},
\end{equation}
where the momentum fractions $x_{1,2}$ satisfy $x_{1}+x_{2}=1$.
With these internal variables in hand, one can define some useful
quantities for both incoming and outgoing mesons, for example: 
\begin{align}
M_{0}^{\prime2}=(e'_{1}+e_{2})^{2}=\frac{p'{}_{\perp}^{2}+m'{}_{1}^{2}}{x_{1}}+\frac{p'{}_{\perp}^{2}+m{}_{2}^{2}}{x_{2}},\quad & \widetilde{M}'_{0}=\sqrt{M_{0}^{\prime2}-(m'_{1}-m_{2})^{2}},\nonumber \\
e_{i}^{(\prime)}=\sqrt{m_{i}^{(\prime)2}+p'{}_{\perp}^{2}+p'{}_{z}^{2}}, & \quad p'_{z}=\frac{x_{2}M_{0}^{\prime}}{2}-\frac{m_{2}^{2}+p'{}_{\perp}^{2}}{2x_{2}M'_{0}},
\end{align}
where $M'{}_{0}^{2}$ can be viewed as the kinetic invariant mass
squared of the incoming $q\bar{q}$ system, and $e_{i}^{(\prime)}$
the corresponding quark energy.

\subsection{Decay constants}

The decay constants for the s-wave and p-wave $D_{sJ}$ mesons are defined
by 
\begin{align}
\langle0|A_{\mu}|P(P^{\prime})\rangle\equiv\mathcal{A}_{\mu}^{P}=if_{P}P_{\mu}^{\prime},\quad & \langle0|V_{\mu}|S(P^{\prime})\rangle\equiv\mathcal{A}_{\mu}^{S}=f_{S}P_{\mu}^{\prime},\nonumber \\
\langle0|V_{\mu}|V(P^{\prime},\varepsilon^{\prime})\rangle\equiv\mathcal{A}_{\mu}^{V}=M_{V}^{\prime}f_{V}\varepsilon_{\mu}^{\prime},\quad & \langle0|A_{\mu}|{}^{3(1)}A(P^{\prime},\varepsilon^{\prime})\rangle\equiv\mathcal{A}_{\mu}^{^{3}A(^{1}A)}=M_{^{3}A(^{1}A)}^{\prime}f_{^{3}A(^{1}A)}\varepsilon_{\mu}^{\prime},\label{eq:decay_constants}
\end{align}
where we have denoted $^{2S+1}L_{J}=\,^{1}S_{0},\,^{3}S_{1},\,^{3}P_{0},\,^{3}P_{1},\,^{1}P_{1}$
and $^{3}P_{2}$ meson states by $P,V,S,^{3}A,{}^{1}A$ and $T$,
respectively. Note that the $^{3}P_{2}$ meson state cannot be produced
by a current \cite{Cheng:2003sm}. 

As an example, we consider the matrix element $\mathcal{A}_{\mu}^{P}$
in Eq. (\ref{eq:decay_constants}), which can be readily obtained
as: 
\begin{equation}
\mathcal{A}_{\mu}^{P}=-i^{2}\frac{N_{c}}{(2\pi)^{4}}\int d^{4}p'_{1}\frac{H'_{P}}{N'_{1}N_{2}}s_{\mu}^{P},
\end{equation}
where 
\begin{align}
s_{\mu}^{P} & \equiv{\rm Tr}[\gamma_{\mu}\gamma_{5}(\cancel{p}'_{1}+m'_{1})\gamma_{5}(-\cancel{p}{}_{2}+m{}_{2})],\nonumber \\
N'_{1} & =p'{}_{1}^{2}-m'{}_{1}^{2}+i\epsilon,\quad N_{2}=p_{2}^{2}-m_{2}^{2}+i\epsilon.
\end{align}
By integrating out $p'{}_{1}^{-}$ and taking into account the
contribution of so-called zero mode, we obtain the decay constant
for a pseudoscalar meson \cite{Jaus:1999zv,Cheng:2003sm}: 
\begin{equation}
f_{P}=\frac{N_{c}}{16\pi^{3}}\int dx_{2}d^{2}p'_{\perp}\frac{h'_{P}}{x_{1}x_{2}(M'{}^{2}-M'{}_{0}^{2})}4(m'_{1}x_{2}+m_{2}x_{1}).
\end{equation}
The explicit form of $h'_{P}$ is given by 
\begin{align}
h'_{P}= & (M'{}^{2}-M'{}_{0}^{2})\sqrt{\frac{x_{1}x_{2}}{N_{c}}}\frac{1}{\sqrt{2}\widetilde{M}'_{0}}\varphi',
\label{eq:hp_P}
\end{align}
where $\varphi'$ is the light-front momentum distribution amplitude
for an s-wave meson. In practice, the following Gaussian-type wave
function is usually adopted:
\begin{align}
\varphi'=\varphi'(x_{2},p'_{\perp})= & 4\left(\frac{\pi}{\beta'{}^{2}}\right)^{3/4}\sqrt{\frac{e'_{1}e_{2}}{x_{1}x_{2}M'_{0}}}\exp\left(-\frac{p'{}_{z}^{2}+p'{}_{\perp}^{2}}{2\beta'{}^{2}}\right),\label{eq:sw_momentum_wf}
\end{align}
where the shape parameter $\beta'$ characterizes the momentum distribution
inside the meson and is expected to be of order $\Lambda_{{\rm QCD}}$. 

Similarly, one can also obtain the decay constant expressions for other mesons \cite{Jaus:1999zv,Cheng:2003sm}: 
\begin{align}
f_{S}= & \frac{N_{c}}{16\pi^{3}}\int dx_{2}d^{2}p'_{\perp}\frac{h'_{S}}{x_{1}x_{2}(M'{}^{2}-M'{}_{0}^{2})}4(m'_{1}x_{2}-m_{2}x_{1}),\nonumber \\
f_{V}= & \frac{N_{c}}{4\pi^{3}M'}\int dx_{2}d^{2}p'_{\perp}\frac{h'_{V}}{x_{1}x_{2}(M'{}^{2}-M'{}_{0}^{2})}\nonumber \\
 & \times\left[x_{1}M'{}_{0}^{2}-m'_{1}(m'_{1}-m_{2})-p'{}_{\perp}^{2}+\frac{m'_{1}+m_{2}}{\omega'_{V}}p'{}_{\perp}^{2}\right],\nonumber \\
f_{^{3}A}= & -\frac{N_{c}}{4\pi^{3}M'}\int dx_{2}d^{2}p'_{\perp}\frac{h'_{^{3}A}}{x_{1}x_{2}(M'{}^{2}-M'{}_{0}^{2})}\nonumber \\
 & \times\left[x_{1}M'{}_{0}^{2}-m'_{1}(m'_{1}+m_{2})-p'{}_{\perp}^{2}-\frac{m'_{1}-m_{2}}{\omega'_{^{3}A}}p'{}_{\perp}^{2}\right],\nonumber \\
f_{^{1}A}= & \frac{N_{c}}{4\pi^{3}M'}\int dx_{2}d^{2}p'_{\perp}\frac{h'_{^{1}A}}{x_{1}x_{2}(M'{}^{2}-M'{}_{0}^{2})}(\frac{m'_{1}-m_{2}}{\omega'_{^{1}A}}p'{}_{\perp}^{2}).
\end{align}
The explicit forms of $h_{M}^{\prime}$ and $\omega_{M}^{\prime}$
can be found in Appendix \ref{app:FF}. 

In practice, the decay constant is used to determine the shape parameter
$\beta'$ in the light-front momentum distribution amplitude, see, for example, Eq. (\ref{eq:sw_momentum_wf}). 

\subsection{Form factors}

The semileptonic decays $\bar{B}_{s}\to D_{sJ}l\bar{\nu}$ are induced
at the quark level through $b\to cl\bar{\nu}$, and the corresponding
effective Hamiltonian reads: 
\begin{equation}
\mathcal{H}_{eff}=\frac{G_{F}}{\sqrt{2}}V_{cb}[\bar{c}\gamma_{\mu}(1-\gamma_{5})b][\bar{l}\gamma^{\mu}(1-\gamma_{5})\nu],
\end{equation}
where $G_{F}$ and $V_{cb}$ are the Fermi constant and CKM matrix
element, respectively. The amplitude of $\bar{B}_{s}\to D_{sJ}l\bar{\nu}$
can be factorized into the product of a hadron transition matrix element
and a leptonic bispinor. The latter can be analytically
calculated, while the former is characterized by form factors. The
$\bar{B}_{s}\to D_{s},D_{s}^{*}$ form factors are defined as follows:
\begin{align}
\langle D_{s}(P'')|V_{\mu}|\bar{B}_{s}(P')\rangle & =(P_{\mu}-\frac{m_{\bar{B}_{s}}^{2}-m_{D_{s}}^{2}}{q^{2}})F_{1}^{\bar{B}_{s}D_{s}}(q^{2})+\frac{m_{\bar{B}_{s}}^{2}-m_{D_{s}}^{2}}{q^{2}}q_{\mu}F_{0}^{\bar{B}_{s}D_{s}}(q^{2}),\nonumber \\
\langle D_{s}^{*}(P'',\varepsilon'')|V_{\mu}|\bar{B}_{s}(P')\rangle & =-\frac{1}{m_{\bar{B}_{s}}+m_{D_{s}^{*}}}\epsilon_{\mu\nu\alpha\beta}\varepsilon''{}^{*\nu}P^{\alpha}q^{\beta}V^{\bar{B}_{s}D_{s}^{*}}(q^{2}),\nonumber \\
\langle D_{s}^{*}(P'',\varepsilon'')|A_{\mu}|\bar{B}_{s}(P')\rangle & =2im_{D_{s}^{*}}\frac{\varepsilon''{}^{*}\cdot q}{q^{2}}q_{\mu}A_{0}^{\bar{B}_{s}D_{s}^{*}}(q^{2})+i(m_{\bar{B}_{s}}+m_{D_{s}^{*}})A_{1}^{\bar{B}_{s}D_{s}^{*}}(q^{2})\left[\varepsilon''{}_{\mu}^{*}-\frac{\varepsilon''{}^{*}\cdot q}{q^{2}}q_{\mu}\right]\nonumber \\
 & -i\frac{\varepsilon''{}^{*}\cdot P}{m_{\bar{B}_{s}}+m_{D_{s}^{*}}}A_{2}^{\bar{B}_{s}D_{s}^{*}}(q^{2})\left[P_{\mu}-\frac{m_{\bar{B}_{s}}^{2}-m_{D_{s}^{*}}^{2}}{q^{2}}q_{\mu}\right],\label{eq:FF_P2PV}
\end{align}
where $P=P^{\prime}+P^{\prime\prime}$, $q=P^{\prime}-P^{\prime\prime}$,
and the convention $\epsilon_{0123}=1$ is adopted. 
The $\bar{B}_{s}\to D_{s0},D_{s1}^{(\prime)}$ form factors are defined in a similar way: 
\begin{align}
\langle D_{s0}(P'')|A_{\mu}|\bar{B}_{s}(P')\rangle & =-i\left[(P_{\mu}-\frac{m_{\bar{B}_{s}}^{2}-m_{D_{s0}}^{2}}{q^{2}})F_{1}^{\bar{B}_{s}D_{s0}}(q^{2})+\frac{m_{\bar{B}_{s}}^{2}-m_{D_{s0}}^{2}}{q^{2}}q_{\mu}F_{0}^{\bar{B}_{s}D_{s0}}(q^{2})\right],\nonumber \\
\langle D_{s1}^{(\prime)}(P'',\varepsilon'')|V_{\mu}|\bar{B}_{s}(P')\rangle & =-2m_{D_{s1}^{(\prime)}}\frac{\varepsilon''{}^{*}\cdot q}{q^{2}}q_{\mu}V_{0}^{\bar{B}_{s}D_{s1}^{(\prime)}}(q^{2})-(m_{\bar{B}_{s}}+m_{D_{s1}^{(\prime)}})V_{1}^{\bar{B}_{s}D_{s1}^{(\prime)}}(q^{2})\left[\varepsilon''{}_{\mu}^{*}-\frac{\varepsilon''{}^{*}\cdot q}{q^{2}}q_{\mu}\right]\nonumber \\
 & +\frac{\varepsilon''{}^{*}\cdot P}{m_{\bar{B}_{s}}+m_{D_{s1}^{(\prime)}}}V_{2}^{\bar{B}_{s}D_{s1}^{(\prime)}}(q^{2})\left[P_{\mu}-\frac{m_{\bar{B}_{s}}^{2}-m_{D_{s1}^{(\prime)}}^{2}}{q^{2}}q_{\mu}\right],\nonumber \\
\langle D_{s1}^{(\prime)}(P'',\varepsilon'')|A_{\mu}|\bar{B}_{s}(P')\rangle & =-i\frac{1}{m_{\bar{B}_{s}}-m_{D_{s1}^{(\prime)}}}\epsilon_{\mu\nu\alpha\beta}\varepsilon''{}^{*\nu}P^{\alpha}q^{\beta}A^{\bar{B}_{s}D_{s1}^{(\prime)}}(q^{2}).
\end{align}
In analogy with those of $\bar{B}_{s}\to D_{s}^{*}$, the form factors
of $\bar{B}_{s}\to D_{s2}$ are defined by
\begin{align}
\langle D_{s2}(P'',\varepsilon'')|V_{\mu}|\bar{B}_{s}(P')\rangle & =-\frac{2V^{\bar{B}_{s}D_{s2}}(q^{2})}{m_{\bar{B}_{s}}+m_{D_{s2}}}\epsilon^{\mu\nu\rho\sigma}(\varepsilon_{T}^{*})_{\nu}(P')_{\rho}(P'')_{\sigma},\nonumber \\
\langle D_{s2}(P'',\varepsilon'')|A_{\mu}|\bar{B}_{s}(P')\rangle & =2im_{D_{s2}}\frac{\varepsilon_{T}^{*}\cdot q}{q^{2}}q_{\mu}A_{0}^{\bar{B}_{s}D_{s2}}(q^{2})+i(m_{\bar{B}_{s}}+m_{D_{s2}})A_{1}^{\bar{B}_{s}D_{s2}}(q^{2})\left[\varepsilon_{T}^{*}{}_{\mu}-\frac{\varepsilon{}_{T}^{*}\cdot q}{q^{2}}q_{\mu}\right]\nonumber \\
 & -i\frac{\varepsilon_{T}^{*}\cdot q}{m_{\bar{B}_{s}}+m_{D_{s2}}}A_{2}^{\bar{B}_{s}D_{s2}}(q^{2})\left[P_{\mu}-\frac{m_{\bar{B}_{s}}^{2}-m_{D_{s2}}^{2}}{q^{2}}q_{\mu}\right],
\end{align}
with
\begin{equation}
\varepsilon_{T\mu}(h)=\frac{1}{m_{\bar{B}_{s}}}\varepsilon''_{\mu\nu}(h)P'{}^{\nu}.
\end{equation}
The explicit form of spin-2 polarization tensor $\varepsilon''_{\mu\nu}$
can be found, for example, in Ref. \cite{Shi:2016gqt}.

In what follows, we will consider how to compute these hadron transition
matrix elements in covariant LFQM. As an example, we consider the
$P\to P$ transition 
\begin{equation}
\langle P(P'')|V_{\mu}|P(P')\rangle\equiv\mathcal{B}_{\mu}^{PP}.
\end{equation}
It is straightforward to obtain 
\begin{equation}
\mathcal{B}_{\mu}^{PP}=-i^{3}\frac{N_{c}}{(2\pi)^{4}}\int d^{4}p'_{1}\frac{H'_{P}H''_{P}}{N'_{1}N''_{1}N_{2}}S_{V\mu}^{PP},
\end{equation}
where 
\begin{align}
S_{V\mu}^{PP} & ={\rm Tr}[\gamma_{5}(\cancel{p}''_{1}+m''_{1})\gamma_{\mu}(\cancel{p}'{}_{1}+m'{}_{1})\gamma_{5}(-\cancel{p}_{2}+m_{2})],\nonumber \\
N_{1}^{\prime\prime} & =p_{1}^{\prime\prime2}-m_{1}^{\prime\prime2}+i\epsilon.
\end{align}
Here we consider the $q^{+}=0$ frame. By integrating out $p'{}_{1}^{-}$
and taking into account the zero mode contribution, one can obtain
the form factors $F_{1}^{PP}$ and $F_{0}^{PP}$ in Eq. (\ref{eq:FF_P2PV}) \cite{Jaus:1999zv,Cheng:2003sm}: 
\begin{equation}
F_{1}^{PP}(q^{2})=f_{+}(q^{2}),\quad F_{0}^{PP}(q^{2})=f_{+}(q^{2})+\frac{q^{2}}{q\cdot P}f_{-}(q^{2}),\label{eq:FF_PP}
\end{equation}
where 
\begin{align}
f_{+}(q^{2})= & \frac{N_{c}}{16\pi^{3}}\int dx_{2}d^{2}p'_{\perp}\frac{h'_{P}h''_{P}}{x_{2}\hat{N}'_{1}\hat{N}''_{1}}\left[x_{1}(M'{}_{0}^{2}+M''{}_{0}^{2})+x_{2}q^{2}-x_{2}(m'_{1}-m''_{1})^{2}-x_{1}(m'_{1}-m_{2})^{2}-x_{1}(m''_{1}-m_{2})^{2}\right],\nonumber \\
f_{-}(q^{2})= & \frac{N_{c}}{16\pi^{3}}\int dx_{2}d^{2}p'_{\perp}\frac{2h'_{P}h''_{P}}{x_{2}\hat{N}'_{1}\hat{N}''_{1}}\Big\{-x_{1}x_{2}M'{}^{2}-p'{}_{\perp}^{2}-m'_{1}m_{2}+(m''_{1}-m_{2})\nonumber \\
 & \times(x_{2}m'_{1}+x_{1}m_{2})+2\frac{q\cdot P}{q^{2}}\left(p'{}_{\perp}^{2}+2\frac{(p'_{\perp}\cdot q_{\perp})^{2}}{q^{2}}\right)+2\frac{(p'_{\perp}\cdot q_{\perp})^{2}}{q^{2}}\nonumber \\
 & -\frac{p'_{\perp}\cdot q_{\perp}}{q^{2}}\left[M''{}^{2}-x_{2}(q^{2}+q\cdot P)-(x_{2}-x_{1})M'{}^{2}+2x_{1}M'{}_{0}^{2}-2(m'_{1}-m_{2})(m'_{1}+m''_{1})\right]\Big\}.
\end{align}
In the above equation, $\hat{N}_{1}^{\prime(\prime\prime)}=x_{1}(M^{\prime(\prime\prime)2}-M_{0}^{\prime(\prime\prime)2})$,
and $h^{\prime(\prime\prime)}_{P}$ can be found in Eq. (\ref{eq:hp_P}).
The expressions for other form factors can be found in Appendix \ref{app:FF}. 

Before closing this subsection, there is a subtle point worth pointing out. 
In the heavy quark limit, the heavy quark spin decouples with the total angular momentum of 
the light degree of freedom. The latter, at this point, is a good quantum number, which can
be used to label the two p-wave axial-vector states as $P_{1}^{3/2}$
and $P_{1}^{1/2}$. These two axial-vector states are mixtures of
$^{3}P_{1}$ and $^{1}P_{1}$ \cite{Cheng:2003sm}:
\begin{equation}
|P_{1}^{3/2}\rangle=\sqrt{\frac{2}{3}}|^{1}P_{1}\rangle+\sqrt{\frac{1}{3}}|^{3}P_{1}\rangle,\quad|P_{1}^{1/2}\rangle=\sqrt{\frac{1}{3}}|^{1}P_{1}\rangle-\sqrt{\frac{2}{3}}|^{3}P_{1}\rangle.\label{eq:mixing}
\end{equation}
In this work, we approximate the two axial-vector states $D_{s1}(2460)$
and $D_{s1}(2536)$ that have been discovered experimentally as eigenstates
of $P_{1}^{1/2}$ and $P_{1}^{3/2}$, respectively.
In the following, we denote them as $D_{s1}$ and $D_{s1}^{\prime}$ for short.
The corresponding form factors can be easily obtained via Eq. (\ref{eq:mixing}).

\section{Numerical results of form factors and phenomenological applications}

\subsection{Input parameters}

In this work, the following constituent quark masses are adopted (in
units of GeV): 
\begin{equation}
m_{s}=0.37,\quad m_{c}=1.4,\quad m_{b}=4.64,
\end{equation}
which are exactly the same as those in Ref. \cite{Cheng:2003sm}.
Similar parameters can also be found in Refs. 
\cite{Lu:2007sg,Wang:2007sxa,Wang:2008xt,Wang:2008ci,Wang:2009mi,Chen:2009qk,Li:2010bb}.

The masses of $\bar{B}_{s}$ and $D_{sJ}$ are taken from PDG (in
units of GeV) \cite{ParticleDataGroup:2024cfk}: 
\begin{align}
 & m_{\bar{B}_{s}}=5.367,\quad m_{D_{s}}=1.968,\quad m_{D_{s}^{*}}=2.112,\nonumber \\
 & m_{D_{s0}}=2.318,\quad m_{D_{s1}}=2.460,\quad m_{D'_{s1}}=2.535,\quad m_{D_{s2}}=2.569.\label{eq:massBsDsJ}
\end{align}

The shape parameter $\beta$, which characterizes the momentum distribution
inside the meson, is determined by fitting the decay constant. With
$f_{\bar{B}_{s}}=230.3\ {\rm MeV}$ from a recent review by FLAG \cite{FlavourLatticeAveragingGroupFLAG:2024oxs},
$f_{D_{s}}=249.9\ {\rm MeV}$ from PDG \cite{ParticleDataGroup:2024cfk},
and $f_{D_{s}^{*}}=272\ {\rm MeV}$ and $f_{D_{s0}}=74.4\ {\rm MeV}$
from Ref. \cite{Verma:2011yw}, one can determine the shape parameters as follows:
\begin{align}
 & \beta_{\bar{B}_{s}}=0.6209\ {\rm GeV},\quad\beta_{D_{s}}=0.5416\ {\rm GeV},\quad\beta_{D_{s}^{*}}=0.4364\ {\rm GeV},\nonumber \\
 & \beta_{D_{s0}}=\beta_{D_{s1}}=\beta_{D_{s1}^{\prime}}=\beta_{D_{s2}}=0.3483\ {\rm GeV},
 \label{eq:beta}
\end{align}
where we have assumed that the shape parameters for $D_{s1}$, $D_{s1}^{\prime}$
and $D_{s2}$ are approximately equal to that of $D_{s0}$.

\subsection{Form factors}

\begin{table}
\caption{The $\bar{B}_{s}\to D_{s},D_{s}^{*},D_{s0},D_{s1},D'_{s1}$,$D_{s2}$
form factors in covariant LFQM, which are fitted using Eq. (\ref{eq:fit_formula_1})
except for $F_{0}^{\bar{B}_{s}D_{s0}}, A^{\bar{B}_{s}D_{s1}}, V_{0,1,2}^{\bar{B}_{s}D_{s1}}$,
$A^{\bar{B}_{s}D'_{s1}}$ and $V_{1}^{\bar{B}_{s}D_{s1}^{\prime}}$.
For the latter cases, the formula in Eq. (\ref{eq:fit_formula_2})
is adopted.}
\centering{}\label{Tab:FF}%
\begin{tabular}{c|c|c|c|c|c|c|c}
\hline 
$F$ & $F(0)$ & $m_{{\rm fit}}$ & $\delta$ & $F$ & $F(0)$ & $m_{{\rm fit}}$ & $\delta$\tabularnewline
\hline 
$F_{1}^{\bar{B}_{s}D_{s}}$ & 0.707 & $4.809$ & $0.259$ & $F_{0}^{\bar{B}_{s}D_{s}}$ & 0.707 & $6.501$ & $0.00874$\tabularnewline
\hline 
$V^{\bar{B}_{s}D_{s}^{*}}$ & 0.813 & $4.774$ & $0.293$ & $A_{0}^{\bar{B}_{s}D_{s}^{*}}$ & 0.693 & $4.526$ & $0.353$\tabularnewline
\hline 
$A_{1}^{\bar{B}_{s}D_{s}^{*}}$ & 0.684 & $6.719$ & $0.142$ & $A_{2}^{\bar{B}_{s}D_{s}^{*}}$ & 0.672 & $4.924$ & $0.289$\tabularnewline
\hline 
$F_{1}^{\bar{B}_{s}D_{s0}}$ & 0.199 & $6.681$ & $0.869$ & $F_{0}^{\bar{B}_{s}D_{s0}}$ & 0.199 & $3.361$ & $0.919$\tabularnewline
\hline 
$A^{\bar{B}_{s}D_{s1}}$ & -0.0772 & $1.008$ & $0.703$ & $V_{0}^{\bar{B}_{s}D_{s1}}$ & 0.0252 & $2.005$ & $0.285$\tabularnewline
\hline 
$V_{1}^{\bar{B}_{s}D_{s1}}$ & -0.0838 & $2.842$ & $0.471$ & $V_{2}^{\bar{B}_{s}D_{s1}}$ & -0.0636 & $5.728$ & $2.269$\tabularnewline
\hline 
$A^{\bar{B}_{s}D'_{s1}}$ & 0.191 & $3.373$ & $0.749$ & $V_{0}^{\bar{B}_{s}D'_{s1}}$ & 0.382 & $3.670$ & $1.195$\tabularnewline
\hline 
$V_{1}^{\bar{B}_{s}D'_{s1}}$ & 0.523 & $9.490$ & $5.059$ & $V_{2}^{\bar{B}_{s}D'_{s1}}$ & -0.0564 & $3.347$ & $2.349$\tabularnewline
\hline 
$V^{\bar{B}_{s}D_{s2}}$ & -0.494 & $4.583$ & $0.472$ & $A_{0}^{\bar{B}_{s}D_{s2}}$ & 0.459 & $3.704$ & $0.526$\tabularnewline
\hline 
$A_{1}^{\bar{B}_{s}D_{s2}}$ & 0.143 & $3.214$ & $2.909$ & $A_{2}^{\bar{B}_{s}D_{s2}}$ & -0.438 & $4.535$ & $0.424$\tabularnewline
\hline 
\end{tabular}
\end{table}

With the inputs given in previous subsection, we calculate the $\bar{B}_{s}\to D_{s},D_{s}^{*},D_{s0},D_{s1},D'_{s1}$
and $D_{s2}$ form factors in covariant LFQM. In order to access the
$q^{2}$ dependence, we adopt the following fit formula: 
\begin{equation}
F(q^{2})=\frac{F(0)}{1-\frac{q^{2}}{m_{{\rm fit}}^{2}}+\delta\left(\frac{q^{2}}{m_{{\rm fit}}^{2}}\right)^{2}}.\label{eq:fit_formula_1}
\end{equation}
However, for $F_{0}^{\bar{B}_{s}D_{s0}}, A^{\bar{B}_{s}D_{s1}}, V_{0,1,2}^{\bar{B}_{s}D_{s1}}$,
$A^{\bar{B}_{s}D'_{s1}}$ and $V_{1}^{\bar{B}_{s}D_{s1}^{\prime}}$, we find that the fitted results for
$m_{{\rm fit}}^{2}$ are negative, and thus we use the following revised
formula: 
\begin{equation}
F(q^{2})=\frac{F(0)}{1+\frac{q^{2}}{m_{{\rm fit}}^{2}}+\delta\left(\frac{q^{2}}{m_{{\rm fit}}^{2}}\right)^{2}}.\label{eq:fit_formula_2}
\end{equation}
Our fitted results for $F(0)$, $m_{{\rm fit}}$ and $\delta$ are
presented in Table \ref{Tab:FF}.

In Ref. \cite{Cheng:2003sm}, the authors considered the $B\to D_{J}$
transitions. The form factors of $\bar{B}_{s}\to D_{sJ}$ should be
close to those of $B\to D_{J}$ if the heavy quark limit is considered.
For comparison, we list the $B\to D_{J}$ form factors in Ref. \cite{Cheng:2003sm} as follows: 
\begin{align}
 & F_{1}^{BD}(0)=0.67,\quad F_{0}^{BD}(0)=0.67,\nonumber \\
 & V^{BD^{*}}(0)=0.75,\quad A_{0}^{BD^{*}}(0)=0.64,\nonumber \\
 & A_{1}^{BD^{*}}(0)=0.63,\quad A_{2}^{BD^{*}}(0)=0.61,\nonumber \\
 & F_{1}^{BD_{0}}(0)=0.24,\quad F_{0}^{BD_{0}}(0)=0.24,\nonumber \\
 & A^{BD_{1}^{1/2}}(0)=-0.12,\quad V_{0}^{BD_{1}^{1/2}}(0)=0.08,\nonumber \\
 & V_{1}^{BD_{1}^{1/2}}(0)=-0.19,\quad V_{2}^{BD_{1}^{1/2}}(0)=-0.12,\nonumber \\
 & A^{BD_{1}^{3/2}}(0)=0.23,\quad V_{0}^{BD_{1}^{3/2}}(0)=0.47,\nonumber \\
 & V_{1}^{BD_{1}^{3/2}}(0)=0.55,\quad V_{2}^{BD_{1}^{3/2}}(0)=-0.09.
\end{align}
As expected, these form factors are indeed very close to the $\bar{B}_{s}\to D_{sJ}$
form factors we are considering.

\begin{figure}
\begin{center}
\includegraphics[scale=1.2]{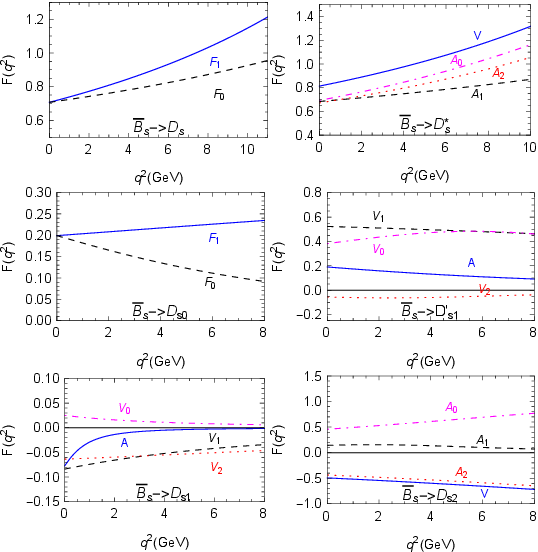}
\caption{
The $q^2$-dependence of the $\bar{B}_{s}\to D_{sJ}$ form factors, where the fit formula in 
Eq. (\ref{eq:fit_formula_1}) or Eq. (\ref{eq:fit_formula_2}) is adopted.
}
\label{fig:ff_Bs2DsJ}
\end{center}
\end{figure}

The $q^{2}$ dependence of the $\bar{B}_{s}\to D_{sJ}$ form factors
are shown in Fig. \ref{fig:ff_Bs2DsJ}, from which, one can see that all the form factors
are rather stable against the variation of $q^{2}$. 

\subsection{Semileptonic $\bar{B}_{s}$ decays}

The decay amplitude of $\bar{B}_{s}\to D_{sJ}l\bar{\nu}$ can be divided
into Lorentz-invariant hadronic part and leptonic part, so that the
decay width can be readily evaluated. The differential decay width
for $\bar{B}_{s}\to D_{s}l\bar{\nu}$ and $\bar{B}_{s}\to D_{s}^{*}l\bar{\nu}$
turn out to be 
\begin{align}
\frac{d\Gamma(\bar{B}_{s}\to D_{s}l\bar{\nu})}{dq^{2}}= & (1-\hat{m}_{l}^{2})^{2}\frac{\sqrt{\lambda(m_{B_{s}}^{2},m_{D_{s}}^{2},q^{2})}G_{F}^{2}|V_{{\rm CKM}}|^{2}}{384m_{B_{s}}^{3}\pi^{3}}\Big\{(\hat{m}_{l}^{2}+2)\lambda(m_{B_{s}}^{2},m_{D_{s}}^{2},q^{2})F_{1}^{2}(q^{2})\nonumber \\
 & +3\hat{m}_{l}^{2}(m_{B_{s}}^{2}-m_{D_{s}}^{2})^{2}F_{0}^{2}(q^{2})\Big\},\\
\frac{d\Gamma_{L}(\bar{B}_{s}\to D_{s}^{*}l\bar{\nu})}{dq^{2}}= & (1-\hat{m}_{l}^{2})^{2}\frac{\sqrt{\lambda(m_{B_{s}}^{2},m_{D_{s}^{*}}^{2},q^{2})}G_{F}^{2}|V_{{\rm CKM}}|^{2}}{384m_{B_{s}}^{3}\pi^{3}}\Big\{3\hat{m}_{l}^{2}\lambda(m_{B_{s}}^{2},m_{D_{s}^{*}}^{2},q^{2})A_{0}^{2}(q^{2})+(\hat{m}_{l}^{2}+2)\nonumber \\
 & \times\Big|\frac{1}{2m_{D_{s}^{*}}}\left[(m_{B_{s}}^{2}-m_{D_{s}^{*}}^{2}-q^{2})(m_{B_{s}}+m_{D_{s}^{*}})A_{1}(q^{2})-\frac{\lambda(m_{B_{s}}^{2},m_{D_{s}^{*}}^{2},q^{2})}{m_{B_{s}}+m_{D_{s}^{*}}}A_{2}(q^{2})\right]\Big|^{2}\Big\},\label{eq:dGLdq2}\\
\frac{d\Gamma^{\pm}(\bar{B}_{s}\to D_{s}^{*}l\bar{\nu})}{dq^{2}}= & (1-\hat{m}_{l}^{2})^{2}\frac{\sqrt{\lambda(m_{B_{s}}^{2},m_{D_{s}^{*}}^{2},q^{2})}G_{F}^{2}|V_{{\rm CKM}}|^{2}}{384m_{B_{s}}^{3}\pi^{3}}\Big\{(m_{l}^{2}+2q^{2})\lambda(m_{B_{s}}^{2},m_{D_{s}^{*}}^{2},q^{2})\nonumber \\
 & \times\Big|\frac{V(q^{2})}{m_{B_{s}}+m_{D_{s}^{*}}}\mp\frac{(m_{B_{s}}+m_{D_{s}^{*}})A_{1}(q^{2})}{\sqrt{\lambda(m_{B_{s}}^{2},m_{D_{s}^{*}}^{2},q^{2})}}\Big|^{2}\Big\},\label{eq:dGpmdq2}
\end{align}
where $\lambda(m_{B_{s}}^{2},m_{i}^{2},q^{2})=(m_{B_{s}}^{2}+m_{i}^{2}-q^{2})^{2}-4m_{B_{s}}^{2}m_{i}^{2}$
with $i=D_{s},D_{s}^{*}$, and $\hat{m}_{l}=m_{l}/\sqrt{q^{2}}$.
The combined transverse differential decay width and total differential
decay width are given by 
\begin{equation}
\frac{d\Gamma_{T}}{dq^{2}}=\frac{d\Gamma_{+}}{dq^{2}}+\frac{d\Gamma_{-}}{dq^{2}},\quad\frac{d\Gamma}{dq^{2}}=\frac{d\Gamma_{L}}{dq^{2}}+\frac{d\Gamma_{T}}{dq^{2}}.
\end{equation}

The differential decay widths for $\bar{B}_{s}\to D_{s0}l\bar{\nu}$
and $\bar{B}_{s}\to D_{s1}^{(\prime)}l\bar{\nu}$ can be obtained
from those of $\bar{B}_{s}\to D_{s}l\bar{\nu}$ and $\bar{B}_{s}\to D_{s}^{*}l\bar{\nu}$
, with the replacements 
\begin{align}
 & m_{D_{s}}\to m_{D_{s0}},\nonumber \\
 & F_{i}^{\bar{B}_{s}D_{s}}(q^{2})\to F_{i}^{\bar{B}_{s}D_{s0}}(q^{2}),\quad i=0,1
\end{align}
and 
\begin{align}
 & m_{B_{s}}+m_{D_{s}^{*}}\to m_{B_{s}}-m_{D_{s1}^{(\prime)}},\nonumber \\
 & V^{\bar{B}_{s}D_{s}^{*}}(q^{2})\to A^{\bar{B}_{s}D_{s1}^{(\prime)}}(q^{2}),\nonumber \\
 & A_{i}^{\bar{B}_{s}D_{s}^{*}}(q^{2})\to V_{i}^{\bar{B}_{s}D_{s1}^{(\prime)}}(q^{2}),\quad i=0,1,2
\end{align}
respectively. The $d\Gamma_{L}/dq^{2}$ and $d\Gamma^{\pm}/dq^{2}$
for $\bar{B}_{s}\to D_{s2}l\bar{\nu}$ are respectively given by Eq.
(\ref{eq:dGLdq2}) multiplied by $(\sqrt{\frac{2}{3}}\frac{|\vec{p}_{T}|}{m_{D_{s2}}})^{2}$
and Eq. (\ref{eq:dGpmdq2}) multiplied by $(\sqrt{\frac{1}{2}}\frac{|\vec{p}_{T}|}{m_{D_{s2}}})^{2}$,
where $\vec{p}_{T}$ denotes the momentum of $D_{s2}$ in the rest
frame of $\bar{B}_{s}$.

The decay width is obtained by 
\begin{equation}
\Gamma=\int_{m_{l}^{2}}^{(m_{\bar{B}_{s}}-m_{D_{sJ}})^{2}}dq^{2}\frac{d\Gamma}{dq^{2}}.
\end{equation}
Our predictions on the branching fractions of semileptonic decays are
listed in Table \ref{Tab:semi}, where our results are also compared
with those in the literature. When arriving at these results, we have
also used the following inputs from PDG \cite{ParticleDataGroup:2024cfk}:
\begin{align}
 & \tau_{\bar{B}_{s}}=1.527\times10^{-12}\ {\rm s},\nonumber \\
 & m_{e}=0.511\ {\rm MeV},\quad m_{\mu}=0.10566\ {\rm GeV},\quad m_{\tau}=1.777\ {\rm GeV},\nonumber \\
 & G_{F}=1.166\times10^{-5}\ {\rm GeV}^{-2},\quad|V_{cb}|=4.18\times10^{-2}.\label{eq:some_inputs}
\end{align}

Some comments are in order. 
\begin{itemize}
\item Roughly speaking, the branching fractions of $\bar{B}_{s}$ to the
s-wave mesons are larger than those of $\bar{B}_{s}$ to the p-wave
mesons, as expected.
\item For the two semileptonic decays $\bar{B}_{s}\to D_{s}\mu\bar{\nu}_{\mu}$
and $\bar{B}_{s}\to D_{s}^{*}\mu\bar{\nu}_{\mu}$, our calculated
results are in good agreement with the experimental data. For the
semileptonic decay $\bar{B}_{s}\to D'_{s1}\mu\bar{\nu}_{\mu}$, our
result is several times larger than the experimental data, however,
our result is close to that in Ref. \cite{Faustov:2012mt}, and all
theoretical predictions are larger than the experimental data. In our opinion, it
is necessary to repeat the measurement in the experiment.
\item As mentioned earlier, considering the heavy quark limit, we approximate
the two axial-vector states $D_{s1}\equiv D_{s1}(2460)$ and $D_{s1}^{\prime}\equiv D_{s1}(2536)$
as eigenstates of $P_{1}^{1/2}$ and $P_{1}^{3/2}$, respectively.
In reality, the physical $D_{s1}(2460)$ and $D_{s1}(2536)$ are of course the mixtures of $P_{1}^{1/2}$
and $P_{1}^{3/2}$. Also considering that the mass of the charm quark is not large enough,
there may exist a relatively large mixing angle.
This issue deserves further study, and it is worth mentioning that we pointed
out in Ref. \cite{Sun:2023noo} that the nature of hadron mixing comes from gluon exchange.
\end{itemize}

\begin{table}
\caption{Our predictions on the branching fractions of semileptonic $\bar{B}_{s}\to D_{s},D_{s}^{*},D_{s0},D_{s1},D_{s1}^{\prime},D_{s2}$
decays, and comparisons with other results in the literature.}
\label{Tab:semi}
\begin{tabular}{c|c|c|c}
\hline 
Channel  & This work  & Other theoretical predictions  & Experiment\tabularnewline
\hline 
$\bar{B}_{s}\to D_{s}e\bar{\nu}_{e}$  & $2.67\times10^{-2}$ & $2.03\times10^{-2}$\cite{Zhang:2021wnv}, $1.84\times10^{-2}$\cite{Hu:2019bdf},
$2.32\times10^{-2}$\cite{Albertus:2014bfa} & -\tabularnewline
 &  & $2.1\times10^{-2}$\cite{Faustov:2012mt}, $1.0\times10^{-2}$\cite{Li:2009wq},
$2.8\sim3.5\times10^{-2}$\cite{Azizi:2008tt}  & \tabularnewline
\hline 
$\bar{B}_{s}\to D_{s}\mu\bar{\nu}_{\mu}$  & $2.66\times10^{-2}$ & $2.32\times10^{-2}$\cite{Albertus:2014bfa}, $1.0\times10^{-2}$\cite{Li:2009wq}  & $(2.31\pm0.21)\times10^{-2}$\tabularnewline
\hline 
$\bar{B}_{s}\to D_{s}\tau\bar{\nu}_{\tau}$  & $8.00\times10^{-3}$ &  $6.3\times10^{-3}$\cite{Hu:2019bdf}, $6.7\times10^{-3}$\cite{Albertus:2014bfa},
$6.2\times10^{-3}$\cite{Faustov:2012mt},  & -\tabularnewline
 &  & $3.3\times10^{-3}$\cite{Li:2009wq}  & \tabularnewline
\hline 
$\bar{B}_{s}\to D_{s}^{*}e\bar{\nu}_{e}$  & $5.16\times10^{-2}$ &  $4.42\times10^{-2}$\cite{Hu:2019bdf}, $6.26\times10^{-2}$\cite{Albertus:2014bfa},
$5.3\times10^{-2}$\cite{Faustov:2012mt},  & -\tabularnewline
 &  & $7.09\times10^{-2}$\cite{Zhang:2010ur}, $1.89\sim6.61\times10^{-2}$\cite{Azizi:2008vt}  & \tabularnewline
\hline 
$\bar{B}_{s}\to D_{s}^{*}\mu\bar{\nu}_{\mu}$  & $5.11\times10^{-2}$ & $6.26\times10^{-2}$\cite{Albertus:2014bfa}  & $(5.2\pm0.5)\times10^{-2}$\tabularnewline
\hline 
$\bar{B}_{s}\to D_{s}^{*}\tau\bar{\nu}_{\tau}$  & $1.18\times10^{-2}$ & $1.20\times10^{-2}$\cite{Hu:2019bdf}, $1.53\times10^{-2}$\cite{Albertus:2014bfa},
$1.3\times10^{-2}$\cite{Faustov:2012mt}  & -\tabularnewline
\hline 
$\bar{B}_{s}\to D_{s0}e\bar{\nu}_{e}$  & $1.19\times10^{-3}$ &  $1.26\times10^{-3}$\cite{Yang:2025zvo}, $0.72\times10^{-3}$\cite{Zuo:2023ksq},
$1.3\times10^{-3}$\cite{Navarra:2015iea},  & -\tabularnewline
 &  & $3.9\times10^{-3}$\cite{Albertus:2014bfa}, $3.6\times10^{-3}$\cite{Faustov:2012mt},
$6.0\times10^{-3}$\cite{Shen:2012mm},  & \tabularnewline
 &  & $2.3\times10^{-3}$\cite{Li:2009wq}, $0.9\sim2.0\times10^{-3}$\cite{Linder:2004vg}  & \tabularnewline
\hline 
$\bar{B}_{s}\to D_{s0}\mu\bar{\nu}_{\mu}$  & $1.18\times10^{-3}$ &  $1.25\times10^{-3}$\cite{Yang:2025zvo}, $0.71\times10^{-3}$\cite{Zuo:2023ksq},
$3.9\times10^{-3}$\cite{Albertus:2014bfa},  & -\tabularnewline
 &  & $6.0\times10^{-3}$\cite{Shen:2012mm}, $2.3\times10^{-3}$\cite{Li:2009wq}  & \tabularnewline
\hline 
$\bar{B}_{s}\to D_{s0}\tau\bar{\nu}_{\tau}$  & $3.04\times10^{-4}$ &  $1.8\times10^{-4}$\cite{Yang:2025zvo}, $0.57\times10^{-4}$\cite{Zuo:2023ksq},
$4\times10^{-4}$\cite{Albertus:2014bfa}, & -\tabularnewline
 &  & $1.9\times10^{-4}$\cite{Faustov:2012mt}, $8.2\times10^{-4}$\cite{Shen:2012mm},
$5.7\times10^{-4}$\cite{Li:2009wq}  & \tabularnewline
\hline 
$\bar{B}_{s}\to D_{s1}e\bar{\nu}_{e}$  & $2.79\times10^{-3}$ & $1.8\times10^{-4}$\cite{Yang:2025zvo}, $6.48\times10^{-4}$\cite{Zuo:2023ksq},
$3.2\times10^{-3}$\cite{Albertus:2014bfa},  & -\tabularnewline
 &  & $1.9\times10^{-3}$\cite{Faustov:2012mt}  & \tabularnewline
\hline 
$\bar{B}_{s}\to D_{s1}\mu\bar{\nu}_{\mu}$  & $2.67\times10^{-3}$ & $1.8\times10^{-4}$\cite{Yang:2025zvo}, $6.42\times10^{-4}$\cite{Zuo:2023ksq},
$3.2\times10^{-3}$\cite{Albertus:2014bfa}, & -\tabularnewline
 &  & $3.5\times10^{-4}$\cite{Li:2010bb}  & \tabularnewline
\hline 
$\bar{B}_{s}\to D_{s1}\tau\bar{\nu}_{\tau}$  & $3.47\times10^{-5}$ &  $2.0\times10^{-5}$\cite{Yang:2025zvo}, $5.4\times10^{-5}$\cite{Zuo:2023ksq},
$3.0\times10^{-4}$\cite{Albertus:2014bfa}, & -\tabularnewline
 &  & $1.5\times10^{-4}$\cite{Faustov:2012mt}, $9.9\times10^{-6}$\cite{Li:2010bb}  & \tabularnewline
\hline 
$\bar{B}_{s}\to D'_{s1}e\bar{\nu}_{e}$  & $11.4\times10^{-3}$ &  $6.17\times10^{-3}$\cite{Yang:2025zvo}, $6.31\times10^{-3}$\cite{Zuo:2023ksq},
$4.7\times10^{-3}$\cite{Albertus:2014bfa},  & -\tabularnewline
 &  & $8.4\times10^{-3}$\cite{Faustov:2012mt}, $4.9\times10^{-3}$\cite{Aliev:2006gk},
$0.8\sim1.0\times10^{-3}$\cite{Linder:2004vg} & \tabularnewline
\hline 
$\bar{B}_{s}\to D'_{s1}\mu\bar{\nu}_{\mu}$  & $11.1\times10^{-3}$ &  $6.13\times10^{-3}$\cite{Yang:2025zvo}, $6.25\times10^{-3}$\cite{Zuo:2023ksq},
$4.7\times10^{-3}$\cite{Albertus:2014bfa}, & $(2.7\pm0.7)\times10^{-3}$\tabularnewline
 &  & $4.0\times10^{-3}$\cite{Li:2010bb}, $4.9\times10^{-3}$\cite{Aliev:2006gk}  & \tabularnewline
\hline 
$\bar{B}_{s}\to D'_{s1}\tau\bar{\nu}_{\tau}$  & $5.18\times10^{-4}$ &  $6.3\times10^{-4}$\cite{Yang:2025zvo}, $3.8\times10^{-4}$\cite{Zuo:2023ksq},
$4.0\times10^{-4}$\cite{Albertus:2014bfa},  & -\tabularnewline
 &  & $4.9\times10^{-4}$\cite{Faustov:2012mt}, $9.7\times10^{-5}$\cite{Li:2010bb}  & \tabularnewline
\hline 
$\bar{B}_{s}\to D_{s2}e\bar{\nu}_{e}$  & $5.91\times10^{-3}$ & $3.77\times10^{-3}$\cite{Zuo:2023ksq}, $4.4\times10^{-3}$\cite{Albertus:2014bfa},
$6.7\times10^{-3}$\cite{Faustov:2012mt}  & -\tabularnewline
\hline 
$\bar{B}_{s}\to D_{s2}\mu\bar{\nu}_{\mu}$  & $5.76\times10^{-3}$ & $3.73\times10^{-3}$\cite{Zuo:2023ksq}, $4.4\times10^{-3}$\cite{Albertus:2014bfa}  & -\tabularnewline
\hline 
$\bar{B}_{s}\to D_{s2}\tau\bar{\nu}_{\tau}$  & $2.62\times10^{-4}$ & $2.4\times10^{-4}$\cite{Zuo:2023ksq}, $3.0\times10^{-4}$\cite{Albertus:2014bfa},
$2.9\times10^{-4}$\cite{Faustov:2012mt}  & -\tabularnewline
\hline 
\end{tabular}
\end{table}

\subsection{Nonleptonic $\bar{B}_{s}$ decays}

In this subsection, we investigate the two-body nonleptonic $\bar{B}_{s}$
decays within the framework of naive factorization. The effective
Hamiltonian governing $b\to cd\bar{u}$ at the tree level
reads:
\begin{align}
\mathcal{H}_{{\rm eff}}(b\to cd\bar{u})= & \frac{G_{F}}{\sqrt{2}}V_{cb}V_{ud}^{*}\{C_{1}\left[\bar{c}_{\alpha}\gamma^{\mu}(1-\gamma_{5})b_{\beta}\right]\left[\bar{d}_{\beta}\gamma_{\mu}(1-\gamma_{5})u_{\alpha}\right]\nonumber \\
 & +C_{2}\left[\bar{c}_{\alpha}\gamma^{\mu}(1-\gamma_{5})b_{\alpha}\right]\left[\bar{d}_{\beta}\gamma_{\mu}(1-\gamma_{5})u_{\beta}\right]\},
\end{align}
where $C_{1,2}$ are the Wilson coefficients, and $\alpha,\beta$ denote
the color indices.

Using the assumption of naive factorization, one can easily arrive at the
following decay amplitudes:
\begin{align}
i\mathcal{M}(\bar{B}_{s}\to D_{s}\pi)= & Nm_{\bar{B}_{s}}^{2}(1-r_{D_{s}}^{2})F_{0}^{\bar{B}_{s}D_{s}}(m_{\pi}^{2}),\\
i\mathcal{M}(\bar{B}_{s}\to D_{s}^{*}\pi)= & (-i)N\sqrt{\lambda(m_{\bar{B}_{s}}^{2},m_{D_{s}^{*}}^{2},m_{\pi}^{2})}A_{0}^{\bar{B}_{s}D_{s}^{*}}(m_{\pi}^{2}),\\
i\mathcal{M}(\bar{B}_{s}\to D_{s0}\pi)= & (-i)Nm_{\bar{B}_{s}}^{2}(1-r_{D_{s0}}^{2})F_{0}^{\bar{B}_{s}D_{s0}}(m_{\pi}^{2}),\\
i\mathcal{M}(\bar{B}_{s}\to D_{s1}^{(\prime)}\pi)= & (-i)N\sqrt{\lambda(m_{\bar{B}_{s}}^{2},m_{D_{s1}^{(\prime)}}^{2},m_{\pi}^{2})}V_{0}^{\bar{B}_{s}D_{s1}^{(\prime)}}(m_{\pi}^{2}),\\
i\mathcal{M}(\bar{B}_{s}\to D_{s2}\pi)= & (-i)\frac{1}{\sqrt{6}}N\frac{\lambda(m_{\bar{B}_{s}}^{2},m_{D_{s2}}^{2},m_{\pi}^{2})}{m_{\bar{B}_{s}}^{2}r_{D_{s2}}}A_{0}^{\bar{B}_{s}D_{s2}}(m_{\pi}^{2}),
\end{align}
where $r_{D_{sJ}}\equiv m_{D_{sJ}}/m_{\bar{B}_{s}}$, and $N=(G_{F}/\sqrt{2})V_{cb}V_{ud}^{*}a_{1}f_{\pi}$,
with $a_{1}=C_{2}+C_{1}/N_{c}$. 

The decay width for $\bar{B}_{s}\to D_{sJ}\pi$ is given by
\begin{equation}
\Gamma=\frac{|\vec{p}_{1}|}{8\pi m_{\bar{B}_{s}}^{2}}|\mathcal{M}|^{2},
\end{equation}
where $|\vec{p}_{1}|$ is the magnitude of the three-momentum of $D_{sJ}$
in the rest frame of $\bar{B}_{s}$. 

Our predictions on the branching fractions of nonleptonic decays are
listed in Table \ref{Tab:non}, where our results are also compared
with those in the literature. When arriving at these results, we have also
used the following inputs \cite{ParticleDataGroup:2024cfk,Shi:2016gqt,Chen:2011ut,Faustov:2012mt}:
\begin{align}
m_{\pi} & =139.57039\ {\rm MeV},\quad m_{K}=493.677\ {\rm MeV},\nonumber \\
f_{\pi} & =130.4\ {\rm MeV},\quad f_{K}=156\ {\rm MeV},\nonumber \\
|V_{ud}| & =0.97435,\quad|V_{us}|=0.22501,\quad a_{1}=1.02.
\end{align}

Some comments are in order. 
\begin{itemize}
\item As shown in Table \ref{Tab:non}, our results for these four decay
channels $\bar{B}_{s}\to D_{s}^{(*)}+\pi/K$ are larger than the experimental
data. Perhaps this indicates a significant deviation from the assumption
of naive factorization, considering our predictions on the corresponding semileptonic
decays are in good agreement with the experimental data.
\item At present, there is still a lack of experimental data on the nonleptonic
$\bar{B}_{s}$ to p-wave $D_{sJ}$ decays. The branching
fractions of $\bar{B}_{s}\to D_{sJ}\pi$ with $J=0,1^{\prime},2$ are sizable,
so these three pionic decays are promising to be discovered experimentally.
\item Compared with other theoretical predictions, most of ours for the nonleptonic
$\bar{B}_{s}$ to p-wave $D_{sJ}$ decays are underestimated,
which may be due to our shape parameter for p-wave $D_{sJ}$ mesons being set too small, as shown
in Eq. (\ref{eq:beta}). For these two decays $\bar{B}_{s}\to D{}_{s1}+\pi/K$,
our predictions on the branching fractions are
particularly smaller due to their small form factors, as shown in
Table \ref{Tab:FF}. However, it is worth emphasizing again that our form factors for 
$\bar{B}_{s}\to D_{sJ}$ are close to those of $B\to D_{J}$ in the classic literature \cite{Cheng:2003sm}.
\end{itemize}

\begin{table}
\caption{Our predictions on the branching fractions of nonleptonic $\bar{B}_{s}\to D_{s},D_{s}^{*},D_{s0},D_{s1},D_{s1}^{\prime},D_{s2}$
decays, and comparisons with other results in the literature. }
\label{Tab:non}
\begin{tabular}{c|c|c|c}
\hline 
Channel & This work & Other theoretical predictions & Experiment\tabularnewline
\hline 
$\bar{B}_{s}\to D_{s}\pi$ & $4.62\times10^{-3}$ & $3.5\times10^{-3}$\cite{Faustov:2012mt}, $2.7\times10^{-3}$\cite{Chen:2011ut},
$1.7\times10^{-3}$\cite{Li:2009wq}, & $(2.98\pm0.14)\times10^{-3}$\tabularnewline
 &  & $2.13\times10^{-3}$\cite{Li:2008ts} & \tabularnewline
\hline 
$\bar{B}_{s}\to D_{s}K$ & $3.52\times10^{-4}$ & $2.8\times10^{-4}$\cite{Faustov:2012mt}, $2.1\times10^{-4}$\cite{Chen:2011ut},
$1.3\times10^{-4}$\cite{Li:2009wq}, & $(2.25\pm0.12)\times10^{-4}$\tabularnewline
 &  & $1.71\times10^{-4}$\cite{Li:2008ts} & \tabularnewline
\hline 
$\bar{B}_{s}\to D_{s}^{*}\pi$ & $4.12\times10^{-3}$ & $2.7\times10^{-3}$\cite{Faustov:2012mt}, $3.0\times10^{-3}$\cite{Chen:2011ut},
$2.8\times10^{-3}$\cite{Li:2010bb}, & $(1.9_{-0.4}^{+0.5})\times10^{-3}$\tabularnewline
 &  & $2.11\times10^{-3}$\cite{Azizi:2008ty} & \tabularnewline
\hline 
$\bar{B}_{s}\to D_{s}^{*}K$ & $3.09\times10^{-4}$ & $2.1\times10^{-4}$\cite{Faustov:2012mt}, $2.4\times10^{-4}$\cite{Chen:2011ut},
$4.8\times10^{-4}$\cite{Li:2010bb} & $(1.32_{-0.32}^{+0.40})\times10^{-4}$\tabularnewline
\hline 
$\bar{B}_{s}\to D_{s0}\pi$ & $3.02\times10^{-4}$ & $1.0\times10^{-3}$\cite{Albertus:2014bfa}, $9\times10^{-4}$\cite{Faustov:2012mt},
$5.2\times10^{-4}$\cite{Li:2009wq} & -\tabularnewline
\hline 
$\bar{B}_{s}\to D_{s0}K$ & $2.38\times10^{-5}$ & $9\times10^{-5}$\cite{Albertus:2014bfa}, $1.2\times10^{-4}$\cite{Faustov:2012mt},
$4\times10^{-5}$\cite{Li:2009wq} & -\tabularnewline
\hline 
$\bar{B}_{s}\to D{}_{s1}\pi$ & $4.39\times10^{-6}$ & $7\times10^{-4}$\cite{Albertus:2014bfa}, $2.9\times10^{-4}$\cite{Faustov:2012mt},
$3.0\times10^{-5}$\cite{Li:2010bb} & -\tabularnewline
\hline 
$\bar{B}_{s}\to D{}_{s1}K$ & $2.87\times10^{-7}$ & $5.4\times10^{-5}$\cite{Albertus:2014bfa}, $2.1\times10^{-5}$\cite{Faustov:2012mt},
$2.3\times10^{-6}$\cite{Li:2010bb} & -\tabularnewline
\hline 
$\bar{B}_{s}\to D_{s1}^{\prime}\pi$ & $0.974\times10^{-3}$ & $1.5\times10^{-3}$\cite{Albertus:2014bfa}, $1.9\times10^{-3}$\cite{Faustov:2012mt},
$1.5\times10^{-3}$\cite{Li:2010bb} & -\tabularnewline
\hline 
$\bar{B}_{s}\to D_{s1}^{\prime}K$ & $7.32\times10^{-5}$ & $1.2\times10^{-4}$\cite{Albertus:2014bfa}, $1.4\times10^{-4}$\cite{Faustov:2012mt},
$1.2\times10^{-4}$\cite{Li:2010bb} & -\tabularnewline
\hline 
$\bar{B}_{s}\to D_{s2}\pi$ & $1.36\times10^{-3}$ & $1.0\times10^{-3}$\cite{Albertus:2014bfa}, $1.6\times10^{-3}$\cite{Faustov:2012mt} & -\tabularnewline
\hline 
$\bar{B}_{s}\to D_{s2}K$ & $9.58\times10^{-5}$ & $8.0\times10^{-5}$\cite{Albertus:2014bfa}, $1.2\times10^{-4}$\cite{Faustov:2012mt} & -\tabularnewline
\hline 
\end{tabular}
\end{table}

\section{Conclusions}

In this work, we investigate semileptonic and nonleptonic $\bar{B}_{s}\to D_{sJ}$ decays
under the framework of covariant light-front quark model (LFQM).
The $\bar{B}_{s}\to D_{sJ}$ transitions are induced at the quark level
by $b\to c$ process, with a spectator anti-strange quark.
We first calculate the transition form factors under the framework of LFQM,
and then apply the derived form factors to analyze the
corresponding semileptonic and nonleptonic decays.
We find that, some branching fractions are sizable, particularly for 
the pionic decays $\bar{B}_{s}\to D_{sJ}\pi$ with $J=0,1^{\prime},2$,
making them promising candidates for observation at the
LHC and future $e^{+}$-$e^{-}$ colliders. 

In conclusion, the $\bar{B}_{s}\to D_{sJ}$ transitions provide a
rich testing ground for exploring the non-perturbative QCD dynamics,
and a feasible solution for studying the internal structure of $D_{sJ}$ mesons.
Our application of covariant LFQM to the $\bar{B}_{s}\to D_{sJ}$
transitions further validates in heavy-light systems. 
Our work is expected to be of great value for establishing corresponding decay channels,
and also be helpful in understanding the non-perturbative QCD dynamics. 

\section*{Acknowledgements}

This work is supported in part by National Natural Science Foundation
of China under Grants No. 12465018.

\appendix
\section{EXPLICIT EXPRESSIONS FOR FORM FACTORS}
\label{app:FF}

For the sake of completeness, in this Appendix, we provide analytical expressions for all the form factors.
If readers want to learn more details, they can refer to Refs. \cite{Jaus:1999zv,Cheng:2003sm}.

For the $P\to P$ transition, the form factors have been
given in the main text, see Eq. (\ref{eq:FF_PP}).

The analytic expressions for the $P\to V$ transition form factors
are given by
\begin{align}
 & V^{PV}(q^{2})=-(m_{P}+m_{V})g(q^{2}),\quad A_{1}^{PV}(q^{2})=-\frac{f(q^{2})}{m_{P}+m_{V}},\quad A_{2}^{PV}(q^{2})=(m_{P}+m_{V})a_{+}(q^{2}),\nonumber \\
 & A_{0}^{B_{c}V}(q^{2})=\frac{m_{P}+m_{V}}{2m_{V}}A_{1}(q^{2})-\frac{m_{P}-m_{V}}{2m_{V}}A_{2}(q^{2})-\frac{q^{2}}{2m_{V}}a_{-}(q^{2}),
\end{align}
where the auxiliary functions $g$, $f$, $a_{+}$ and $a_{-}$ are:
\begin{eqnarray}
g(q^{2}) & = & -\frac{N_{c}}{16\pi^{3}}\int dx_{2}d^{2}p_{\perp}^{\prime}\frac{2h_{P}^{\prime}h_{V}^{\prime\prime}}{x_{2}\hat{N}_{1}^{\prime}\hat{N}_{1}^{\prime\prime}}\Bigg\{ x_{2}m_{1}^{\prime}+x_{1}m_{2}+(m_{1}^{\prime}-m_{1}^{\prime\prime})\frac{p_{\perp}^{\prime}\cdot q_{\perp}}{q^{2}}+\frac{2}{w_{V}^{\prime\prime}}\Bigg[p_{\perp}^{\prime2}+\frac{(p_{\perp}^{\prime}\cdot q_{\perp})^{2}}{q^{2}}\Bigg]\Bigg\},\nonumber \\
f(q^{2}) & = & \frac{N_{c}}{16\pi^{3}}\int dx_{2}d^{2}p_{\perp}^{\prime}\frac{h_{P}^{\prime}h_{V}^{\prime\prime}}{x_{2}\hat{N}_{1}^{\prime}\hat{N}_{1}^{\prime\prime}}\Bigg\{2x_{1}(m_{2}-m_{1}^{\prime})(M_{0}^{\prime2}+M_{0}^{\prime\prime2})-4x_{1}m_{1}^{\prime\prime}M_{0}^{\prime2}\nonumber \\
 &  & +2x_{2}m_{1}^{\prime}q\cdot P+2m_{2}q^{2}-2x_{1}m_{2}(M^{\prime2}+M^{\prime\prime2})+2(m_{1}^{\prime}-m_{2})(m_{1}^{\prime}+m_{1}^{\prime\prime})^{2}\nonumber \\
 &  & +8(m_{1}^{\prime}-m_{2})\Bigg[p_{\perp}^{\prime2}+\frac{(p_{\perp}^{\prime}\cdot q_{\perp})^{2}}{q^{2}}\Bigg]+2(m_{1}^{\prime}+m_{1}^{\prime\prime})(q^{2}+q\cdot P)\frac{p_{\perp}^{\prime}\cdot q_{\perp}}{q^{2}}\nonumber \\
 &  & -4\frac{q^{2}p_{\perp}^{\prime2}+(p_{\perp}^{\prime}\cdot q_{\perp})^{2}}{q^{2}w_{V}^{\prime\prime}}\Bigg[2x_{1}(M^{\prime2}+M_{0}^{\prime2})-q^{2}-q\cdot P-2(q^{2}+q\cdot P)\frac{p_{\perp}^{\prime}\cdot q_{\perp}}{q^{2}}\nonumber \\
 &  & -2(m_{1}^{\prime}-m_{1}^{\prime\prime})(m_{1}^{\prime}-m_{2})\Bigg]\Bigg\},\nonumber \\
a_{+}(q^{2}) & = & \frac{N_{c}}{16\pi^{3}}\int dx_{2}d^{2}p_{\perp}^{\prime}\frac{2h_{P}^{\prime}h_{V}^{\prime\prime}}{x_{2}\hat{N}_{1}^{\prime}\hat{N}_{1}^{\prime\prime}}\Bigg\{(x_{1}-x_{2})(x_{2}m_{1}^{\prime}+x_{1}m_{2})-[2x_{1}m_{2}+m_{1}^{\prime\prime}+(x_{2}-x_{1})m_{1}^{\prime}]\frac{p_{\perp}^{\prime}\cdot q_{\perp}}{q^{2}}\nonumber \\
 &  & -2\frac{x_{2}q^{2}+p_{\perp}^{\prime}\cdot q_{\perp}}{x_{2}q^{2}w_{V}^{\prime\prime}}[p_{\perp}^{\prime}\cdot p_{\perp}^{\prime\prime}+(x_{1}m_{2}+x_{2}m_{1}^{\prime})(x_{1}m_{2}-x_{2}m_{1}^{\prime\prime})]\Bigg\},\nonumber \\
a_{-}(q^{2}) & = & \frac{N_{c}}{16\pi^{3}}\int dx_{2}d^{2}p_{\perp}^{\prime}\frac{h_{P}^{\prime}h_{V}^{\prime\prime}}{x_{2}\hat{N}_{1}^{\prime}\hat{N}_{1}^{\prime\prime}}\Bigg\{2(2x_{1}-3)(x_{2}m_{1}^{\prime}+x_{1}m_{2})-8(m_{1}^{\prime}-m_{2})\Bigg[\frac{p_{\perp}^{\prime2}}{q^{2}}+2\frac{(p_{\perp}^{\prime}\cdot q_{\perp})^{2}}{q^{4}}\Bigg]\nonumber \\
 &  & -[(14-12x_{1})m_{1}^{\prime}-2m_{1}^{\prime\prime}-(8-12x_{1})m_{2}]\frac{p_{\perp}^{\prime}\cdot q_{\perp}}{q^{2}}\nonumber \\
 &  & +\frac{4}{w_{V}^{\prime\prime}}\Bigg([M^{\prime2}+M^{\prime\prime2}-q^{2}+2(m_{1}^{\prime}-m_{1}^{\prime\prime})(m_{1}^{\prime}-m_{2})](A_{3}^{(2)}+A_{4}^{(2)}-A_{2}^{(1)})+Z_{2}(3A_{2}^{(1)}-2A_{4}^{(2)}-1)\nonumber \\
 &  & +\frac{1}{2}[x_{1}(q^{2}+q\cdot P)-2M^{\prime2}-2p_{\perp}^{\prime}\cdot q_{\perp}-2m_{1}^{\prime}(m_{1}^{\prime\prime}+m_{2})-2m_{2}(m_{1}^{\prime}-m_{2})](A_{1}^{(1)}+A_{2}^{(1)}-1)\nonumber \\
 &  & +q\cdot P\Bigg[\frac{p_{\perp}^{\prime2}}{q^{2}}+\frac{(p_{\perp}^{\prime}\cdot q_{\perp})^{2}}{q^{4}}\Bigg](4A_{2}^{(1)}-3)\Bigg)\Bigg\}.
\end{eqnarray}

For $P\to S$ and $P\to A$ transitions:
\begin{align}
 & F_{1}^{PS}(q^{2})=-u_{+}(q^{2}),\quad F_{0}^{PS}(q^{2})=-u_{+}(q^{2})-\frac{q^{2}}{q\cdot P}u_{-}(q^{2}),\nonumber \\
 & A^{PA}(q^{2})=-(m_{P}-m_{A})q(q^{2}),\quad V_{1}^{PA}(q^{2})=-\frac{\ell(q^{2})}{m_{P}-m_{A}},\quad V_{2}^{PA}(q^{2})=(m_{P}-m_{A})c_{+}(q^{2}),\nonumber \\
 & V_{0}^{PA}(q^{2})=\frac{m_{P}-m_{A}}{2m_{T}}V_{1}(q^{2})-\frac{m_{P}+m_{A}}{2m_{A}}V_{2}(q^{2})-\frac{q^{2}}{2m_{A}}c_{-}(q^{2}),
\end{align}
where $u_{\pm}$ and $\ell,q,c_{\pm}$ can be readily obtained by
making the following replacements 
\begin{eqnarray}
u_{\pm}(q^{2}) & = & -f_{\pm}(q^{2})|_{m_{1}^{\prime\prime}\to-m_{1}^{\prime\prime},\;h_{P}^{\prime\prime}\to h_{S}^{\prime\prime}},\nonumber \\{}
[\ell^{\,^{3}A,\,^{1}A}(q^{2}),q^{\,^{3}A,\,^{1}A}(q^{2}),c_{\pm}^{\,^{3}A,\,^{1}A}(q^{2})] & = & [f(q^{2}),g(q^{2}),a_{\pm}(q^{2})]|_{m_{1}^{\prime\prime}\to-m_{1}^{\prime\prime},\;h_{V}^{\prime\prime}\to h_{\,^{3}A,\,^{1}A}^{\prime\prime},\;w_{V}^{\prime\prime}\to w_{\,^{3}A,\,^{1}A}^{\prime\prime}},
\end{eqnarray}
where only the $1/W^{\prime\prime}$ terms in the $P\to\,^{1}A$ form
factors are kept. It should be cautious that the replacement of $m_{1}^{\prime\prime}\to-m_{1}^{\prime\prime}$
should not be applied to $w^{\prime\prime}$
and $h^{\prime\prime}$. 

The $P\to T$ transition form factors are 
\begin{align}
 & V^{PT}(q^{2})=-m_{P}(m_{P}+m_{T})h(q^{2}),\quad A_{1}^{PT}(q^{2})=-\frac{m_{P}}{m_{P}+m_{T}}k(q^{2}),\quad A_{2}^{PT}(q^{2})=m_{P}(m_{P}+m_{T})b_{+}(q^{2}),\nonumber \\
 & A_{0}^{PT}(q^{2})=\frac{m_{P}+m_{T}}{2m_{T}}A_{1}(q^{2})-\frac{m_{P}-m_{T}}{2m_{T}}A_{2}(q^{2})-\frac{m_{P}q^{2}}{2m_{T}}b_{-}(q^{2}),
\end{align}
where
\begin{eqnarray}
h(q^{2}) & = & -g(q^{2})|_{h_{V}^{\prime\prime}\to h_{T}^{\prime\prime}}+\frac{N_{c}}{16\pi^{3}}\int dx_{2}d^{2}p_{\perp}^{\prime}\frac{2h_{P}^{\prime}h_{T}^{\prime\prime}}{x_{2}\hat{N}_{1}^{\prime}\hat{N}_{1}^{\prime\prime}}\Bigg[(m_{1}^{\prime}-m_{1}^{\prime\prime})(A_{3}^{(2)}+A_{4}^{(2)})\nonumber \\
 &  & +(m_{1}^{\prime\prime}+m_{1}^{\prime}-2m_{2})(A_{2}^{(2)}+A_{3}^{(2)})-m_{1}^{\prime}(A_{1}^{(1)}+A_{2}^{(1)})+\frac{2}{w_{V}^{\prime\prime}}(2A_{1}^{(3)}+2A_{2}^{(3)}-A_{1}^{(2)})\Bigg],\nonumber \\
k(q^{2}) & = & -f(q^{2})|_{h_{V}^{\prime\prime}\to h_{T}^{\prime\prime}}+\frac{N_{c}}{16\pi^{3}}\int dx_{2}d^{2}p_{\perp}^{\prime}\frac{h_{P}^{\prime}h_{T}^{\prime\prime}}{x_{2}\hat{N}_{1}^{\prime}\hat{N}_{1}^{\prime\prime}}\Bigg\{2(A_{1}^{(1)}+A_{2}^{(1)})\nonumber \\
 &  & \times[m_{2}(q^{2}-\hat{N}_{1}^{\prime}-\hat{N}_{1}^{\prime\prime}-m_{1}^{\prime2}-m_{1}^{\prime\prime2})-m_{1}^{\prime}(M^{\prime\prime2}-\hat{N}_{1}^{\prime\prime}-m_{1}^{\prime\prime2}-m_{2}^{2})\nonumber \\
 &  & -m_{1}^{\prime\prime}(M^{\prime2}-\hat{N}_{1}^{\prime}-m_{1}^{\prime2}-m_{2}^{2})-2m_{1}^{\prime}m_{1}^{\prime\prime}m_{2}]+2(m_{1}^{\prime}+m_{1}^{\prime\prime})\Bigg(A_{2}^{(1)}Z_{2}+\frac{q\cdot P}{q^{2}}A_{1}^{(2)}\Bigg)\nonumber \\
 &  & +16(m_{2}-m_{1}^{\prime})(A_{1}^{(3)}+A_{2}^{(3)})+4(2m_{1}^{\prime}-m_{1}^{\prime\prime}-m_{2})A_{1}^{(2)}\nonumber \\
 &  & +\frac{4}{w_{V}^{\prime\prime}}\Bigg([M^{\prime2}+M^{\prime\prime2}-q^{2}+2(m_{1}^{\prime}-m_{2})(m_{1}^{\prime\prime}+m_{2})](2A_{1}^{(3)}+2A_{2}^{(3)}-A_{1}^{(2)})\nonumber \\
 &  & -4\Bigg[A_{2}^{(3)}Z_{2}+\frac{q\cdot P}{3q^{2}}(A_{1}^{(2)})^{2}\Bigg]+2A_{1}^{(2)}Z_{2}\Bigg)\Bigg\},\nonumber \\
b_{+}(q^{2}) & = & -a_{+}(q^{2})|_{h_{V}^{\prime\prime}\to h_{T}^{\prime\prime}}+\frac{N_{c}}{16\pi^{3}}\int dx_{2}d^{2}p_{\perp}^{\prime}\frac{h_{P}^{\prime}h_{T}^{\prime\prime}}{x_{2}\hat{N}_{1}^{\prime}\hat{N}_{1}^{\prime\prime}}\Bigg\{8(m_{2}-m_{1}^{\prime})(A_{3}^{(3)}+2A_{4}^{(3)}+A_{5}^{(3)})\nonumber \\
 &  & -2m_{1}^{\prime}(A_{1}^{(1)}+A_{2}^{(1)})(A_{2}^{(2)}+A_{3}^{(2)})+2(m_{1}^{\prime}+m_{1}^{\prime\prime})(A_{2}^{(2)}+2A_{3}^{(2)}+A_{4}^{(2)})\nonumber \\
 &  & +\frac{2}{w_{V}^{\prime\prime}}[2[M^{\prime2}+M^{\prime\prime2}-q^{2}+2(m_{1}^{\prime}-m_{2})(m_{1}^{\prime\prime}+m_{2})](A_{3}^{(3)}+2A_{4}^{(3)}+A_{5}^{(3)}-A_{2}^{(2)}-A_{3}^{(2)})\nonumber \\
 &  & +[q^{2}-\hat{N}_{1}^{\prime}-\hat{N}_{1}^{\prime\prime}-(m_{1}^{\prime}+m_{1}^{\prime\prime})^{2}](A_{2}^{(2)}+2A_{3}^{(2)}+A_{4}^{(2)}-A_{1}^{(1)}-A_{2}^{(1)})]\Bigg\},\nonumber \\
b_{-}(q^{2}) & = & -a_{-}(q^{2})|_{h_{V}^{\prime\prime}\to h_{T}^{\prime\prime}}+\frac{N_{c}}{16\pi^{3}}\int dx_{2}d^{2}p_{\perp}^{\prime}\frac{h_{P}^{\prime}h_{T}^{\prime\prime}}{x_{2}\hat{N}_{1}^{\prime}\hat{N}_{1}^{\prime\prime}}\Bigg\{8(m_{2}-m_{1}^{\prime})(A_{4}^{(3)}+2A_{5}^{(3)}+A_{6}^{(3)})\nonumber \\
 &  & -6m_{1}^{\prime}(A_{1}^{(1)}+A_{2}^{(1)})+4(2m_{1}^{\prime}-m_{1}^{\prime\prime}-m_{2})(A_{3}^{(2)}+A_{4}^{(2)})\nonumber \\
 &  & +2(3m_{1}^{\prime}+m_{1}^{\prime\prime}-2m_{2})(A_{2}^{(2)}+2A_{3}^{(2)}+A_{4}^{(2)})\nonumber \\
 &  & +\frac{2}{w_{V}^{\prime\prime}}\Bigg[2[M^{\prime2}+M^{\prime\prime2}-q^{2}+2(m_{1}^{\prime}-m_{2})(m_{1}^{\prime\prime}+m_{2})]\nonumber \\
 &  & \times(A_{4}^{(3)}+2A_{5}^{(3)}+A_{6}^{(3)}-A_{3}^{(2)}-A_{4}^{(2)})+2Z_{2}(3A_{4}^{(2)}-2A_{6}^{(3)}-A_{2}^{(1)})\nonumber \\
 &  & +2\frac{q\cdot P}{q^{2}}\Bigg(6A_{2}^{(1)}A_{1}^{(2)}-6A_{2}^{(1)}A_{2}^{(3)}+\frac{2}{q^{2}}(A_{1}^{(2)})^{2}-A_{1}^{(2)}\Bigg)\nonumber \\
 &  & +[q^{2}-2M^{\prime2}+\hat{N}_{1}^{\prime}-\hat{N}_{1}^{\prime\prime}-(m_{1}^{\prime}+m_{1}^{\prime\prime})^{2}+2(m_{1}^{\prime}-m_{2})^{2}](A_{2}^{(2)}+2A_{3}^{(2)}+A_{4}^{(2)}-A_{1}^{(1)}-A_{2}^{(1)})\Bigg]\Bigg\}.
\end{eqnarray}

In the above equations, $A_{j}^{(i)}$ and $Z_{2}$ are given by 
\begin{align}
 & A_{1}^{(1)}=\frac{x_{1}}{2},\quad A_{2}^{(1)}=A_{1}^{(1)}-\frac{p_{\perp}^{\prime}\cdot q_{\perp}}{q^{2}},\nonumber \\
 & A_{1}^{(2)}=-p_{\perp}^{\prime2}-\frac{(p_{\perp}^{\prime}\cdot q_{\perp})^{2}}{q^{2}},\;A_{2}^{(2)}=(A_{1}^{(1)})^{2},\;A_{3}^{(2)}=A_{1}^{(1)}A_{2}^{(1)},\;A_{4}^{(2)}=(A_{2}^{(1)})^{2}-\frac{1}{q^{2}}A_{1}^{(2)},\nonumber \\
 & A_{1}^{(3)}=A_{1}^{(1)}A_{1}^{(2)},\,A_{2}^{(3)}=A_{2}^{(1)}A_{1}^{(2)},\,A_{3}^{(3)}=A_{1}^{(1)}A_{2}^{(2)},\,A_{4}^{(3)}=A_{2}^{(1)}A_{2}^{(2)},\,A_{5}^{(3)}=A_{1}^{(1)}A_{4}^{(2)},\,A_{6}^{(3)}=A_{2}^{(1)}A_{4}^{(2)}-\frac{2}{q^{2}}A_{2}^{(1)}A_{1}^{(2)},\nonumber \\
 & Z_{2}=\hat{N}_{1}^{\prime}+m_{1}^{\prime2}-m_{2}^{2}+(1-2x_{1})M^{\prime2}+(q^{2}+q\cdot P)\frac{p_{\perp}^{\prime}\cdot q_{\perp}}{q^{2}},
\end{align}
and the explicit forms of $h_{M}^{\prime}$ and $w_{M}^{\prime}$ are 
\begin{align}
 & h_{P}^{\prime}=h_{V}^{\prime}=(M^{\prime2}-M_{0}^{\prime2})\sqrt{\frac{x_{1}x_{2}}{N_{c}}}\frac{1}{\sqrt{2}\tilde{M}_{0}^{\prime}}\varphi^{\prime},\nonumber \\
 & h_{S}^{\prime}=\sqrt{\frac{2}{3}}h_{\,^{3}A}^{\prime}=(M^{\prime2}-M_{0}^{\prime2})\sqrt{\frac{x_{1}x_{2}}{N_{c}}}\frac{1}{\sqrt{2}\tilde{M}_{0}^{\prime}}\frac{\tilde{M}_{0}^{\prime2}}{2\sqrt{3}M_{0}^{\prime}}\varphi_{p}^{\prime},\nonumber \\
 & h_{\,^{1}A}^{\prime}=h_{T}^{\prime}=(M^{\prime2}-M_{0}^{\prime2})\sqrt{\frac{x_{1}x_{2}}{N_{c}}}\frac{1}{\sqrt{2}\tilde{M}_{0}^{\prime}}\varphi_{p}^{\prime},\nonumber \\
 & w_{V}^{\prime}=M_{0}^{\prime}+m_{1}^{\prime}+m_{2},\quad w_{\,^{3}A}^{\prime}=\frac{\tilde{M}_{0}^{\prime2}}{m_{1}^{\prime}-m_{2}},\quad w_{\,^{1}A}^{\prime}=2,
\end{align}
where $\varphi^{\prime}$ and $\varphi_{p}^{\prime}$ are respectively the light-front
momentum distribution amplitudes for s-wave and p-wave mesons:
\begin{align}
 & \varphi^{\prime}=\varphi^{\prime}(x_{2},p_{\perp}^{\prime})=4\left(\frac{\pi}{\beta^{\prime2}}\right)^{3/4}\sqrt{\frac{dp_{z}^{\prime}}{dx_{2}}}\exp\left(-\frac{p_{z}^{\prime2}+p_{\perp}^{\prime2}}{2\beta^{\prime2}}\right),\quad\frac{dp_{z}^{\prime}}{dx_{2}}=\frac{e_{1}^{\prime}e_{2}}{x_{1}x_{2}M_{0}^{\prime}},\nonumber \\
 & \varphi_{p}^{\prime}=\varphi_{p}^{\prime}(x_{2},p_{\perp}^{\prime})=\sqrt{\frac{2}{\beta^{\prime2}}}\varphi^{\prime}.
\end{align}

\end{document}